\documentstyle [11pt,twoside]{article}
\input{epsf.tex}
   
\begin{document}
\centerline{\Large SMOOTH POTENTIAL CHAOS AND N-BODY SIMULATIONS}
\vskip .2in
\centerline{\large HENRY E. KANDRUP}
\vskip .05in
\centerline{Department of Astronomy, Department of Physics, and}
\centerline{Institute for Fundamental Theory, University of Florida, 
Gainesville, Florida 32611}
\centerline{kandrup@astro.ufl.edu}
\vskip .2in
\centerline{\large IOANNIS V. SIDERIS}
\vskip .05in
\centerline{Department of Astronomy, University of Florida, 
Gainesville, Florida 32611}
\centerline{sideris@astro.ufl.edu}
\vskip .25in
\centerline{ABSTRACT}
\vskip .1in
Integrations in fixed N-body realisations of smooth density distributions 
corresponding to a chaotic galactic potential can
be used to derive reliable estimates of the largest (finite time) 
Lyapunov exponent ${\chi}_{S}$ associated with an orbit in the smooth 
potential generated from the same initial condition, even though the $N$-body 
orbit is typically characterised by an $N$-body exponent 
${\chi}_{N}{\;}{\gg}{\;}{\chi}_{S}$. This can be accomplished either by
comparing initially nearby orbits in a single $N$-body system or by tracking
orbits with the same initial condition evolved in two different $N$-body
realisations of the same smooth density.
\vskip .5in
\centerline{1. INTRODUCTION AND MOTIVATION}
\vskip .1in
At the present time, there are two general approaches which can be used to 
model the structure and evolution of systems like elliptical galaxies. On 
the one hand, one can perform detailed $N$-body simulations, solving the 
coupled equations of motion either exactly or in some approximation.
On the other, one can try to construct equilibrium models as solutions to
the collisionless Boltzmann equation, either analytically ({\em e.g.,} using
the approach developed by Hunter \& Qian [1993]) or numerically ({\em e.g.,}
by implementing some version of Schwarzschild's [1979] method). 

Both approaches
would appear to be extremely useful, but each has its limitations. Despite
significant advances in hardware, direct $N$-body simulations still cannot be 
performed for particle number as large as $N{\;}{\sim}{\;}10^{11}$, so that 
one cannot consider point masses in the integrations as corresponding to 
individual stars in real galaxies. Integrations in
smooth potentials have the advantage that one can consider characteristics
which, presumably, correspond to the orbits of individual stars. However, the
very assumption that the system can be described by a smooth potential 
constitutes an idealisation which, albeit generally accepted, has never been
proven rigorously.

Obvious questions to be answered
thus include
the following: To what extent is it true that, for sufficiently large $N$,
solutions to the full gravitational $N$-body problem can be mimicked by 
motions in a smooth potential? And, to the extent that the smooth potential
approximation is not completely sufficient, to what extent can discreteness
effects really be modeled by friction and noise in the context of a
Fokker-Planck description ({\em cf.} Rosenbluth, MacDonald, \& Judd 1957)?
In particular, to what extent are `real' $N$-body orbits well mimicked by
solutions to a (time-dependent) Langevin equation ({\em cf.} Chandrasekhar
1943) which incorporates dynamical friction and Gaussian white noise? 
Fokker-Planck descriptions were formulated originally to extract
statistical information about long time behaviour, assuming implicitly that
the bulk potential is integrable or nearly integrable. However, recent years
have seen a growing recognition that galactic potentials may admit a fair
amount of chaos; and analyses of the short time behaviour of individual 
orbits in chaotic potentials have provided compelling evidence that friction 
and noise can dramatically accelerate phase space transport (Lieberman \&
Lichtenberg 1972, Lichtenberg \& Wood 1989, Kandrup, Pogorelov, 
\& Sideris 2000, Siopis \& Kandrup 2000). Does this really mean that
discreteness effects can be important already on time scales much shorter
than the relaxation time $t_{R}$?

Closely related to these issues is the nature of the continuum limit. In what
sense is it true that, as $N\to\infty$, orbits in the $N$-body potential
converge towards characteristics in some smooth potential? Superficially, at
least, it might seem that such a convergence is impossible. Dating back to 
Miller (1964), it has been recognised that the $N$-body problem is chaotic
in the sense that individual orbits exhibit exponential sensitivity towards
small changes in initial conditions; and it seems generally accepted today
that, when expressed in units of inverse dynamical times $t_{D}^{-1}$, the 
largest $N$-body Lyapunov exponent ${\chi}_{N}$ does not converge towards 
zero as $N\to\infty$ ({\em cf.} Kandrup \& Smith 1991,
Goodman, Heggie \& Hut 1994), even if the $N$-body system samples an integrable
density distribution. Indeed, recent work, both numerical (Hemsendorf \& 
Merritt 2002) and analytic (Pogorelov 2001), suggests that, even for a density
distribution corresponding to an integrable potential, the largest
Lyapunov exponent may actually {\em increase} with increasing $N$. As probed
in the usual way, the $N$-body problem may become {\em more chaotic} as $N$
increases!

A complete resolution to this apparent conundrum will require long time 
integrations of systems with very large $N$, which is impractical using 
current hardware. However, considerable insight into the continuum limit
can be, and has been, obtained by studying the properties of orbits and orbit 
ensembles evolved in {\em frozen}-$N$ systems, {\em i.e.,} fixed (in time and
space) 
$N$-body realisations of specified smooth density distributions (Kandrup \& 
Sideris 2001, Sideris \& Kandrup 2002). In particular, that work led to several
significant conclusions: (1) The largest $N$-body Lyapunov exponent 
${\chi}_{N}$ does {\em not} decrease with increasing $N$, even for an 
integrable density distribution. However, there is still a clear, 
quantifiable sense in which, 
as $N$ increases, the $N$-body orbits become progressively more similar to
smooth potential characteristics. (2) As $N$ increases the Fourier
spectra associated with $N$-body orbits more closely resemble the spectra
associated with characteristics in the smooth potential, be these either 
regular or chaotic. (3) Alternatively, viewed macroscopically $N$-body orbits
and smooth characteristics with the same initial condition typically diverge
as a power law in time on a time scale $t_{G}(N)$ that increases with 
increasing $N$. For the case of regular characteristics, 
$t_{G}{\;}{\propto}{\;}N^{1/2}t_{D}$; for chaotic characteristics,
$t_{G}{\;}{\propto}{\;}(\ln N)t_{D}$. It follows that, for sufficiently large
$N$, $N$-body orbits and smooth potential characteristics remain close for
comparatively long times. 

This would seem a result of some significance, but it still begs an important
issue. If $N$-body orbits converge towards smooth potential characteristics,
it should be possible, at least for sufficiently large $N$, to extract 
information about any chaos that may be associated with the bulk potential.
In particular, it must be possible to extract estimates of finite time
({\em cf.} Grassberger, Badii, \& Politi 1988)
Lyapunov exponents ${\chi}_{S}$ for motion in the smooth potential, even
though the $N$-body orbits themselves are characterised by exponents 
${\chi}_{N}$ which, typically, are much larger than ${\chi}_{S}$. (Chaotic
orbits in generic smooth potentials typically have a largest Lyapunov
exponent ${\chi}_{S}{\;}{\sim}{\;}t_{D}^{-1}$. For both interacting [Hemsendorf
\& Merritt 2002] and frozen-$N$ [Kandrup \& Sideris 2000] Plummer systems, 
${\chi}_{N}{\;}{\sim}{\;} 20t_{D}^{-1}$ for $N{\;}{\sim}{\;}10^{5}.$)

The aim of this note is to demonstrate that this can in fact be done. In
particular, it will be shown that there are at least two different ways in
which estimates of ${\chi}_{S}$ can be extracted from frozen-$N$ systems,
one involving a comparison of orbits generated from nearby initial conditions
and the other involving integrations of the same initial condition in two
different $N$-body realisations of the same smooth density distribution.
Section 2 describes the algorithms and then exhibits representative results
for two simple model potentials. Section 3 interprets the success of these
algorithms by postulating the existence of two `types' of chaos, 
{\em microscopic chaos}, or {\em microchaos}, which is generic to the $N$-body 
problem, and {\em macroscopic chaos}, or {\em macrochaos}, possibly associated 
with the bulk potential which,
if present, will also be manifested in the $N$-body problem.
\vskip .2in
\centerline{2. NUMERICAL ESTIMATION OF SMOOTH POTENTIAL LYAPUNOV EXPONENTS}
\vskip .1in
\centerline{2.1. {\em Models Considered}}
\vskip .1in
The algorithms described in this Section will be applied to representative
orbits evolved in two different density distributions:
\par\noindent
1. A spherically symmetric Plummer sphere, for which 
\begin{equation}
{\rho}_{P}(r)=\left({3M\over 4{\pi}b^{3}}\right)
\left(1 + {r^{2}\over b^{2}}\right)^{-5/2}.
\end{equation}
This corresponds via Poisson's equation to a spherically symmetric, and
hence integrable, potential 
\begin{equation}
{\Phi}_{P}(r)=-{GM\over \sqrt{r^{2}+b^{2}}}.
\end{equation}
Units were so chosen that $G=M=b=1$. 
\par\noindent
2. A constant density triaxial ellipsoid, for which 
\begin{equation}
{\rho}_{E}({\bf r})={3M\over 4{\pi}abc} \times
\cases {  m^{2} & if $m^{2}{\;}{\le}{\;}1$, \cr
               0     &  if $m^{2} > 1$, \cr}
\end{equation}
with
\begin{equation}
m^{2}=\left( {x^{2}\over a^{2}} + 
{y^{2}\over b^{2}} + {z^{2}\over c^{2}}\right),
\end{equation}
perturbed by a spherically symmetric density spike (black hole) of mass 
$M_{BH}$, this corresponding to a potential 
\begin{equation}
{\Phi}_{E}({\bf r})={\Phi}_{0}+{1\over 2}\left(
{\omega}_{a}^{2}x^{2}+ {\omega}_{b}^{2}y^{2} + {\omega}_{c}^{2}z^{2} \right)
-{GM_{BH}\over \sqrt{r^{2}+{\epsilon}^{2}}},
\end{equation}
with ${\epsilon}=10^{-3}$.
Attention focused on the case $M=1.0$ and  
$M_{BH}=10^{-1.5}M{\;}{\approx}{\;}0.0316228$, and units were again chosen
so that $G=1$.
The axis ratios were taken as
$a=1.95$, $b=1.50$, and $c=1.05$, which yield ({\em cf.} Bertin 2000)
${\Phi}_{0}{\;}{\approx}{\;}-1.00608$,
${\omega}_{a}{\;}{\approx}{\;}0.4663$,
${\omega}_{b}{\;}{\approx}{\;}0.5508$, and 
${\omega}_{c}{\;}{\approx}{\;}0.6753$.
For energies sufficiently small that orbits are restricted to $m<1$, the
phase space is almost completely chaotic (Kandrup \& Sideris 2002). This 
implies that one need not worry about transitions between regular and chaotic 
behaviour which can be induced by discreteness effects in more complex 
potentials like the triaxial generalisations of the Dehnen (1993) potential,
which admit a complex coexistence of both regular and chaotic orbits. 

Frozen-$N$ orbits in these systems were integrated with a particle-particle 
numerical scheme using a variable timestep integrator with accuracy 
parameter $10^{-8}$ which conserved energy to at least one part in $10^{6}$.
The $1/r$ kernels for the individual masses were regulated through the 
introduction of a softening parameter ${\epsilon}=10^{-5}$.
\vskip .1in
\centerline{3.2. {\em The Numerical Algorithms}}
\vskip .1in
\par\noindent
{\em Algorithm 1.} Involving a comparison of orbits generated from nearby 
initial conditions in the same frozen-$N$ density distribution.
\par
The key realisation here is that, even though orbits generated from nearby
initial conditions diverge initially at a rate 
${\Lambda}{\;}{\sim}{\;}{\chi}_{N}$, this divergence quickly saturates
once the growing separation between the orbits becomes large compared with
the typical distance between neighboring point masses. If the smooth potential 
characteristic is regular, this initial exponential divergence is replaced
immediately by a more modest power law divergence which proceeds on a time
scale $t_{G}{\;}{\propto}{\;}N^{1/2}t_{D}$. If instead the characteristic is 
chaotic, the initial exponential divergence at a rate 
${\Lambda}{\;}{\sim}{\;}{\chi}_{N}$ 
is replaced by a slower exponential divergence at a rate 
${\Lambda}{\;}{\sim}{\;}{\chi}_{S}$ which, in turn, typically proceeds until 
the 
separation becomes `macroscopic', {\em i.e.,} comparable to the size of the
entire accessible phase space region. At this point, this second exponential
divergence saturates and is replaced by a power law divergence which proceeds
on a time scale $t_{G}{\;}{\propto}{\;}(\ln N)t_{D}$.

These statements can be corroborated straightforwardly by tracking the 
separation
\begin{equation}
{\delta}r(t)=|{\bf r}_{1}-{\bf r}_{2}|
\end{equation}
and plotting $\ln {\delta}r(t)$ as a function of $t$. By so doing, one 
discovers that, just as for the smooth potential, nearby initial conditions
can exhibit significantly different values of ${\chi}_{S}$. (Nearby initial
conditions evolved in a smooth potential can have significantly different
finite time Lyapunov exponents even if, for late times, these exponents 
converge towards the same asymptotic ${\chi}_{\infty}$.)
If, alternatively, one wishes to estimate a `typical' ${\chi}_{S}$ for orbits 
in some given phase space region, one can select an ensemble of initial
conditions from that region, evolve each initial condition into the future,
and then extract a mean ${\langle}{\chi}_{S}{\rangle}$ from the time-dependent
mean separation
\begin{equation}
{\langle}{\delta}r{\rangle}={1\over k}\sum_{i=1}^{k}{\delta}r_{i}.
\end{equation}

Figure 1 exhibits the results of such a computation for an ensemble of 100
chaotic initial conditions evolved in a frozen-$N$ realisation of the the 
ellipsoid plus black hole system. The  different curves in the Figure 
represent frozen-$N$ backgrounds with $N$ varying between $N=10^{4.5}$ and 
$N=10^{6}$. Figure 2 exhibits analogous data generated for regular initial 
conditions in the Plummer potential. In each case, the initial conditions
sampled a phase space region of size ${\Delta}r{\;}{\sim}{\;}{\Delta}v{\;}
{\sim}{\;}10^{-3}$, with the perturbed orbits being generated from initial 
conditions displaced from the original initial conditions by a distance 
${\delta}r(0)=10^{-5}$ in a randomly chosen direction. 

It is evident that, for both the ellipsoid and Plummer potentials, the mean 
separation ${\langle}{\delta}r{\rangle}$ begins by diverging at a rate that 
is comparable to the value of the largest $N$-body Lyapunov exponent, 
${\chi}_{N}$. For the case of the ellipsoid potential, this exponential 
divergence is (at least for sufficiently large $N$) eventually replaced by a 
slower exponential divergence at a rate ${\sim}{\;}{\chi}_{S}$ which persists 
until ${\delta}r$ becomes `macroscopic,' {\em i.e.,} comparable to the size 
of the accessible phase space region. At this point the exponential divergence 
is replaced by a slower power law separation which proceeds until ${\delta}r$ 
saturates. For the case of the Plummer potential, the second exponential phase 
is absent, the initial exponential phase being replaced immediately by a power 
law growth. This is especially evident from Figure 3, which plots the same 
data as Figure 2 on a linear scale. 

As noted already, for both regular and chaotic potentials the first 
exponential phase typically stops once the 
separation ${\delta}r$ becomes large compared with the mean interparticle 
spacing ${\sim}{\;}n^{-1/3}$, with $n$ a typical number density. For the 
constant density ellipsoid, $n^{-1/3}{\;}{\approx}{\;}0.427N^{-1/3}$. For 
$N=10^{5}$ this corresponds
to $n^{-1/3}{\;}{\approx}{\;}0.00919$ and 
$\ln n^{-1/3}{\;}{\approx}{\;}-4.69$; 
for $N=10^{-6}$, $n^{-1/3}{\;}{\approx}{\;}0.00427$ and 
$\ln n^{-1/3}{\;}{\approx}{\;}-5.45$.
Alternatively, the second, slower exponential phase exhibited by chaotic
potentials typically ceases once 
the separation has become `macroscopic', {\em i.e.,} comparable to the size 
of the accessible phase space region. This macroscopic scale appears to 
coincide with the scale on which two initially nearby orbits evolved in the 
smooth potential will cease their exponential divergence.

As discussed more carefully in Section 3, the comparatively sharp break 
between the first and second phases indicates that one can implement a 
{\em de facto} distinction between {\em microscopic chaos} associated 
primarily with close encounters and {\em macroscopic chaos} associated 
with the bulk potential, even though the effects of these sources of chaos
are not completely decoupled.

It is evident that this prescription for estimating ${\chi}_{S}$ can only work 
for comparatively large values of $N$, where the typical interparticle spacing 
is much smaller than the total size of the accessible configuration space 
region. If this condition is not satisfied, ${\delta}r(t)$ will become 
`macroscopic' almost as soon as it becomes large compared with $n^{-1/3}$, so 
that the intermediate second stage disappears. For the particular model 
exhibited in Figure 1, one requires $N>10^{5}$ or so in order to obtain a 
reasonable estimate of ${\chi}_{S}$. Indeed, it is evident from the dot-dashed
curve in Figure 1 that, after the initial exponential divergence and before
the final saturation, the data for $N=10^{4.5}$ can be well fit by a linear
growth law ${\delta}r=A(t-t_{0})$.
\vskip .1in
\par\noindent
{\em Algorithm 2.} Involving a comparison of orbits generated from a single 
initial condition evolved in two different frozen-$N$ density distributions 
that sample the same smooth density.
\par
The smooth exponent ${\chi}_{S}$ also provides information about the rate of
divergence associated with orbits generated in two different frozen-$N$ 
simulations. Specifically, if one computes ${\delta}r$ for a collection of
orbits evolved in two different frozen-$N$ density distributions associated
with the same chaotic potential, one again finds an evolution manifesting
the same three stages. This is illustrated in Figure 4, which was generated
for the ellipsoid potential for the same initial conditions as Figure 1.
Figure 5 exhibits analogous data generated for the Plummer potential. It is
evident once again that, for the Plummer potential, the intermediate stage is 
absent.

Given that ${\chi}_{S}$ provides information about the rate of divergence
of orbits in two different frozen-$N$ backgrounds, each of which can be viewed
intuitively as a `perturbation' of the smooth density distribution, one might
also expect that ${\chi}_{S}$ provides information about the rate at which
orbits in a single frozen-$N$ simulation diverge from smooth characteristics
with the same initial condition. As illustrated in Figure 6, this expectation
is in fact correct. The macroscopic power law divergence between the frozen-$N$
orbit and the corresponding smooth potential characteristic, which 
parallels the behaviour observed if Algorithm 1 or 2 is implemented, has 
been discussed extensively elsewhere ({\em cf.} Figure 8 in Kandrup \& 
Sideris 2001 and Figure 2 in Sideris \& Kandrup 2002 and the accompanying
discussion).
\vfill\eject

\vskip .1in
\centerline{3.3. {\em Why these algorithms work}}
\vskip .1in
Given the assumption that orbits in an $N$-body system `feel' two different
sorts of chaos which act on different scales, it is not surprising that one
can derive estimates of both the $N$-body Lyapunov exponent ${\chi}_{N}$ and
the smooth potential ${\chi}_{S}$ from a comparison of initially proximate
orbits in a single frozen-$N$ system. However, the fact that ${\chi}_{S}$ is
also related to the rate of divergence of orbits in different frozen-$N$
simulations is, perhaps, less obvious. The key to understanding this 
phenomenon is the fact that discreteness effects really can be mimicked by
dynamical friction and Gaussian white noise in the context of a Langevin 
description.

Specifically (Sideris and Kandrup 2002), at both the level
of individual orbits, as probed, {\em e.g.,} by Fourier spectra, and at the
level of orbit ensembles, as probed, {\em e.g.,} by the efficiency of phase
mixing for both regular and chaotic orbit ensembles, discreteness effects
associated with an $N$-body density distribution are extremely well reproduced
by Gaussian white noise with a `temperature' ${\Theta}{\;}{\sim}{\;}|E|$,
with $E$ the orbital energy, and a coefficient of dynamical friction 
${\eta}{\;}{\propto}{\;}1/N$. This dependence on $N$ is of course very similar
to the scaling $t_{R}{\;}{\sim}{\;}{\eta}^{-1}{\;}{\propto}{\;}N/(\ln N)$ 
predicted in a 
conventional Fokker-Planck description; and, indeed, given the limited range
in $N$ that can be probed numerically (the notion of a smooth potential
appears to break down for $N{\;}{\le}10^{3}$; simulations with 
$N{\;}{\gg}{\;}10^{6}$ become prohibitively expensive computationally!), 
the numerical simulations are completely consistent with this scaling. 

But why does this explain the observed divergence of orbits in different
frozen-$N$ distributions? The crucial point here, as described, {\em e.g.,} 
in Habib, Kandrup \& Mahon (1997) or Kandrup and Novotny (2002), is that, 
viewed mesoscopically, an ensemble
of noisy orbits with fixed ${\Theta}$ and ${\eta}$, each generated from the
same chaotic initial condition or from a set of very nearby initial conditions,
will typically disperse in such as a fashion that
\begin{equation}
{\delta}r{\;}{\propto}{\;}({\Theta}{\eta})^{1/2}\exp({\chi}_{S}t).
\end{equation}
Given the assumed scaling ${\eta}{\;}{\propto}{\;}1/N$, it then follows that
\begin{equation}
\ln {\delta}r = {\rm const} + {1\over 2}\ln {\eta} +{\chi}_{S}t =
{\rm const} -{1\over 2}\ln N + {\chi}_{S}t.
\end{equation}
Numerical simulations demonstrate that noisy orbits diverge at the same rate 
${\chi}_{S}$ observed for orbits in frozen-$N$ simulations; and the connection 
between $N$ and ${\eta}$ implicit in a Fokker-Planck description makes a 
specific prediction as to the $N$-dependence of the exponential prefactor. If 
discreteness effects
really can be modeled as Gaussian noise, ${\delta}r$ should satisfy eq.~(9).
To the extent that the `mean' trajectory associated with the noisy ensemble
coincides, at least approximately, with the smooth potential characteristic,
the same scaling should also be observed when comparing noisy orbits and
smooth potential characteristics. 

This scaling implies (a) that $\ln {\delta}r$ should grow linearly at a rate 
${\chi}_{S}$, independent of $N$, but (b) that, for fixed $t$, an increase in 
$N$ or a decrease in ${\eta}$ by an order of magnitude should decrease 
$\ln {\delta}r$ by $(1/2)\ln 10{\;}{\approx}{\;}1.15$.
That this scaling is in fact realised for both noisy orbits and orbits in
frozen-$N$ backgrounds is evident from Figure 7. Here the solid curves
exhibit the results of noisy integrations of the same initial conditions used 
to generate Figure 1 and 4, all computed with ${\Theta}=1.0$ but allowing for 
values of ${\eta}$ extending from $10^{-4}$ to $10^{-7}$. The dots accompanying
the upper two curves represent results from frozen-$N$ integrations for
$N=10^{4.5}$ and $10^{5.5}$ (the same data plotted in Figure 6). The results 
from Sideris and Kandrup (2002) 
suggest a best fit correspondence $\log_{10}{\eta} = -\log_{10}N + p$, with 
$p{\;}{\approx}{\;}0.5$; and indeed, it is apparent visually that the noisy
and frozen-$N$ curves for, {\em e.g.,} ${\eta}=10^{-5}$ and $N=10^{5.5}$ are 
extremely similar. 

Presuming that this scaling holds for smaller ${\eta}$ as 
well, the lowest curve in Figure 7 should correspond, at least approximately, 
to $N=10^{7.5}$.

\vskip .1in
\centerline{III. DISCUSSION}
\vskip .1in
This paper has described two algorithms which can be used to obtain estimates
of the largest (finite time) Lyapunov exponents ${\chi}_{S}$ associated with 
a smooth density distribution from frozen-$N$ realisations of that density. 
The first 
involves comparing two orbits in a single frozen-$N$ system generated from
nearby initial conditions. The second involves comparing orbits with the same
initial condition evolved in two different frozen-$N$ systems, each sampling 
the
same smooth density distribution. The success of the first algorithm emphasises
the fact that detailed information about the bulk potential really is buried
in an $N$-body simulation. The success of the latter emphasises another 
important point, namely that smooth potential Lyapunov exponents ${\chi}_{S}$
also provide information about the divergence of the same initial condition 
in different $N$-body systems, {\em i.e.,} information about the extent to
which, viewed mesoscopically, discreteness effects limit the intrinsic 
reliability of orbits in a
pointwise sense. As stressed already, the success of this alternative algorithm
reflects the fact that discreteness effects really can be well mimicked by
Gaussian white noise in the context of a Fokker-Planck description.

The key point in all this is that, for the case of a chaotic bulk potential,
two nearby initial conditions will, when evolved into the future, exhibit a
three stage evolution, reflecting the effects of both {\em microscopic} and
{\em macroscopic chaos,} {\em i.e.,} microchaos and macrochaos.
\par\noindent
(1) For early times and small separations, the orbits will diverge 
exponentially at a rate comparable to a typical $N$-body Lyapunov exponent 
${\chi}_{N}$. 
\par\noindent
(2) However, once the separation between the orbits becomes
large compared with the typical interparticle spacing, this divergence 
ceases and is replaced by a slower divergence at a rate 
${\sim}{\;}{\chi}_{S}$, which proceeds until the separation becomes 
macroscopic.
\par\noindent
(3) At still later times the orbits exhibit a more modest power law divergence.

The computations described here only yielded estimates of finite time Lyapunov 
exponents, not the true Lyapunov exponent as defined in a late time limit,
which, given a complex phase space, can be much larger or smaller. This, 
however, is not necessarily bad. Although old in physical time, galaxies are 
young objects when expressed in terms of the dynamical time $t_{D}$ -- 
typically no more than ${\sim}{\;}100-200t_{D}$ in age -- so that
such asympotic limits are not well motivated physically. However, one might
argue that, even though an asymptotic limit is not justified for individual
orbits, the true Lyapunov exponent is important in that ({\em cf.} Kandrup
\& Mahon 1994) it characterises the average instability associated with the
invariant measure, {\em i.e.,} a uniform sampling of the chaotic portions
of the constant energy hypersurface. The obvious point, then, is that to 
obtain an estimate of the true Lyapunov exponent it suffices to repeat the 
calculations described here for an ensemble of initial conditions sampling the
constant energy hypersurface which, as described elsewhere ({\em cf.} Kandrup,
Sideris, \& Bohn 2001) is straightforward numerically. Alternatively, one
can actually compute estimates of ${\chi}_{S}$ in the usual way (Benettin, 
Galgani, and Strelcyn 1976) by tracking the evolution of a small perturbation
which is periodically renormalised, provided only that one makes sure that
the perturbation always remains large compared with the scale on which the
microscopic chaos saturates, {\em i.e.,} very large compared with a typical
interparticle spacing. 

As a practical matter, it would appear that the algorithm can work for any
$N$-body system in which the typical interparticle spacing is sufficiently
small compared to the size of the system. If $N$ is not sufficiently large,
the first exponential phase will not saturate until the separation of the
originally proximate orbits has become macroscopic, so that the second 
exponential phase is lost. For the models considered here, one requires 
$N>10^{5}$ or so. For systems manifesting very high density contrasts, 
{\em e.g.,} triaxial generalisations of the cuspy Dehnen potentials,
one may require much larger $N$ to obtain an adequate sampling of the central
region. In point of fact, however, the requirement of large $N$ is more than
a practical consideration: it would appear that, if $N$ is too small, the very
notion of a bulk potential becomes suspect. One finds, {\em e.g.,} that, for 
the ellipsoid 
plus black hole potential, discreteness effects can be reasonably well 
modeled by Gaussian white noise for $N{\;}{\sim}{\;}10^{4}$, but that this 
model fails for substantially smaller $N$ (Sideris \& Kandrup 2002).

Viewed from the standpoint of nonlinear dynamics, the $N$-body problem --
or at least the frozen-$N$ model considered here -- constitutes an interesting
example of a system
in which chaos can arise for different reasons on different scales.
Because gravity is strong on short scales, one finds generically that 
close encounters between nearby particles trigger chaos on a time scale that
is typically short compared with $t_{D}$. As probed by $N$-body Lyapunov 
exponents computed in the usual way, this microchaos does not decrease 
with increasing $N$; if anything, ${\chi}_{N}$ increases with increasing $N$.
However, the `range' of the chaos, expressed relative to the total size of
the system, {\em does} decrease with increasing $N$ since the typical 
interparticle
spacing scales as $N^{-1/3}$. Alternatively, because of the long range 
character of the gravitational interaction, one also encounters the 
possibility of macrochaos which will arise if the bulk density 
distribution corresponds to a bulk potential that admits global stochasticity. 
Both forms of chaos can play a nontrivial role in the $N$-body problem and,
as has been shown here, information about ${\chi}_{N}$ and ${\chi}_{S}$ can
both be extracted from a judicious analysis of numerical data. 

In this sense, it would appear that although, strictly speaking, the rate at 
which $N$-body orbits diverge is set by the Lyapunov exponent ${\chi}_{N}$, it 
may be misleading to assert (Heggie 1991) that ``the approximation of a smooth 
potential is useful for studying orbits, but not for studying their 
divergence.'' The smooth potential Lyapunov exponent ${\chi}_{S}$ does indeed 
provide useful information regarding the divergence of orbits on mesoscopic 
scales large compared with the interparticle spacing but small compared with 
the size of the system.
\vskip .2in
Portions of this paper were written while HEK was a visitor at the Aspen
Center for Physics, the hospitality of which is acknowledged gratefully.
This work was supported in part by the NSF through AST0070809.
\vskip 0.5in
\centerline{REFERENCES}
\vskip .15in
\par\noindent
Bertin, G. 2000, Dynamics of Galaxies (Cambridge: Cambridge University Press)
\par\noindent
Chandrasekhar, S. 1943, Rev. Mod. Phys. 15, 1
\par\noindent
Dehnen, W. 1993, MNRAS, 265, 250
\par\noindent
Bennetin, G., Galgani, L., \& Strelcyn, J.-M. 1976, Phys. Rev. A 14, 2338
\par\noindent
Goodman, J., Heggie, D. C., \& Hut, P. 1994, ApJ, 415, 715
\par\noindent
Grassberger, P., Badii, R., \& Politi, A. 1988, J. Stat. Phys. 51, 135
\par\noindent
Habib, S., Kandrup, H. E., Mahon, M. E. 1997, ApJ, 480, 155
\par\noindent
Heggie, D. 1991, in Predictability, Stability, and Chaos in $N$-Body Dynamical
Systems, ed. A. \par Roy (New York: Plenum), 47
\par\noindent
Hemsendorf, M. \& Merritt, D. 2002 (astro-ph/0205538)
\par\noindent
Hunter, C. \& Qian, E. 1993, MNRAS, 262, 401 
\par\noindent
Kandrup, H. E. \& Mahon, M. E. 1994, A\&A, 290, 762
\par\noindent
Kandrup, H. E. \& Novotny, S. 2002 (astro-ph/0204019)
\par\noindent
Kandrup, H. E., Pogorelov, I. V., \& Sideris, I. V. 2000, MNRAS, 311, 719
\par\noindent
Kandrup, H. E. \& Sideris, I. V. 2001, Phys. Rev. E 64, 056209
\par\noindent
Kandrup, H. E. \& Sideris, I. V. 2002, Celestial Mechanics 82, 61
\par\noindent
Kandrup, H. E., Bohn, C. L., \& Sideris, I. V. 2001, Phys. Rev. E 65, 016214
\par\noindent
Kandrup, H. E. \& Smith, H. 1991, ApJ, 374, 255
\par\noindent
Lichtenberg, A. J. \& Wood, B. P. 1989, Phys. Rev. Lett. 62, 2213
\par\noindent
Lieberman, M. A. \& Lichtenberg, A. J. 1972, Phys. Rev. A 5, 1852
\par\noindent
Miller, R. H. 1964, ApJ, 140, 250
\par\noindent
Pogorelov, I. V. 2001, University of Florida Ph.~D. dissertation
\par\noindent
Rosenbluth, M. N., MacDonald, W. M., \& Judd, D. L. 1957, Phys. Rev. 107, 1
\par\noindent
Schwarzschild, M. 1979, ApJ, 471, 82
\par\noindent
Sideris, I. V. \& Kandrup, H. E. 2002, Phys. Rev. E 66, 066203
\par\noindent
Siopis, C. \& Kandrup, H. E. 2000, MNRAS, 319, 43
\vfill\eject
\pagestyle{empty}
\begin{figure}[t]
\centering
\centerline{
        \epsfxsize=18cm
        \epsffile{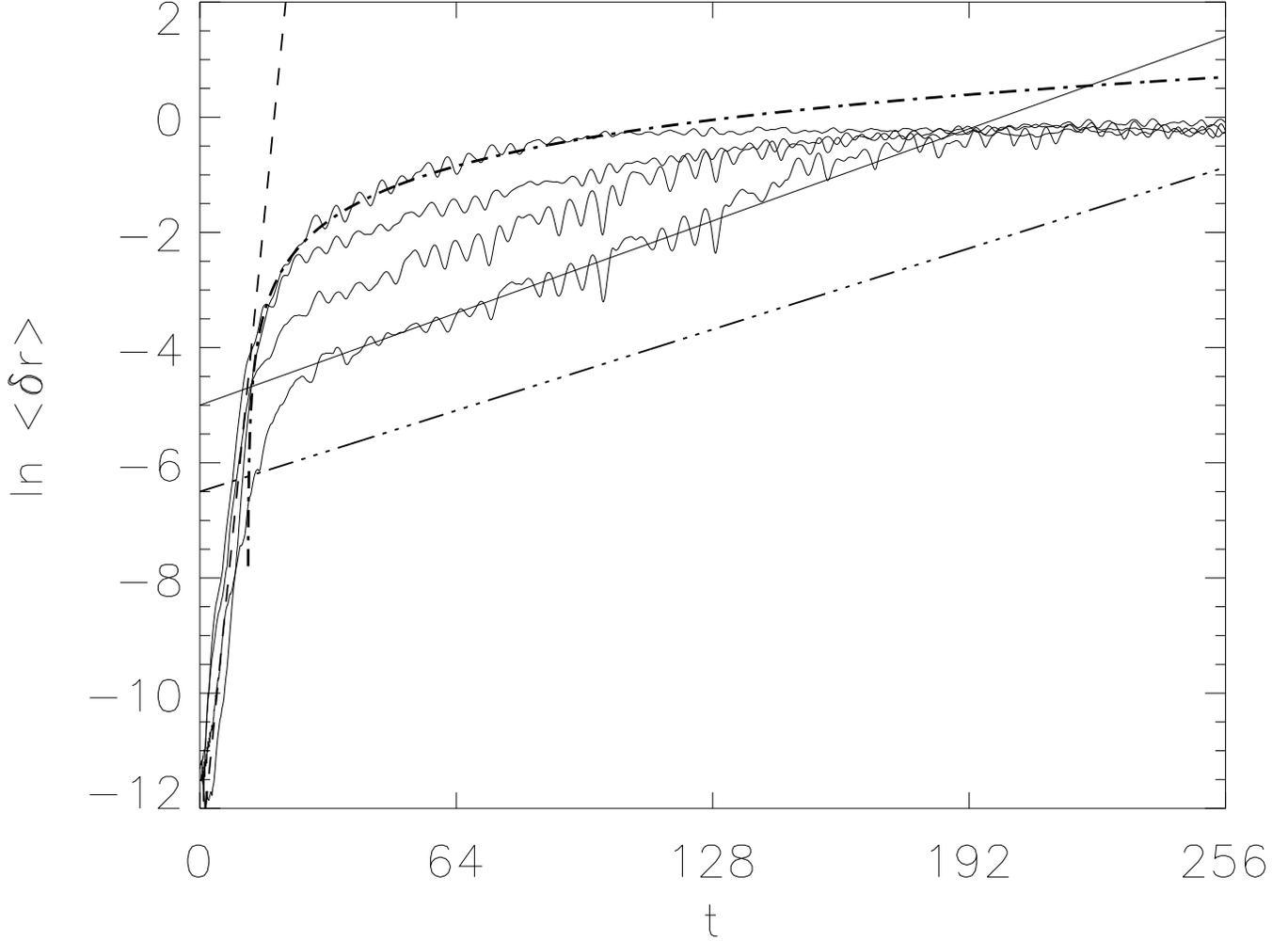}
           }
        \begin{minipage}{18cm}
        \end{minipage}
        \vskip -0.2in\hskip -0.1in
\caption{
The mean spatial separation between orbits generated in frozen-$N$ systems
from initial conditions separated in configuration space by a distance 
${\delta}r(0)=10^{-5}$. Each curve was generated by averaging over 100 pairs
of initial conditions evolved in $N$-body realisations of the chaotic
triaxial ellipsoid plus black hole potential. The four curves correspond 
(from top to bottom) to $N=10^{4.5}$, $10^{5}$, $10^{5.5}$, and $10^{6}$.
The solid line corresponds to a slope of $0.025$, generated as a least 
squares fit to the $N=10^{6}$ data over the interval $32<t<128$.
The triple dot-dashed line has a slope $0.022$, equal to the mean value of the 
smooth potential Lyapunov exponent ${\chi}_{S}$. The dashed line has a slope 
$0.75$, equal to the mean value of the $N$-body Lyapunov exponent ${\chi}_{N}$.
The dot-dashed curve overlaying the data for $N=10^{4.5}$ represents the
function ${\delta}r=A(t-t_{0})$ for $A=0.008$ and $t_{0}=12.0$.
\label{fig1}}
\end{figure}

\pagestyle{empty}
\begin{figure}[t]
\centering
\centerline{
        \epsfxsize=18cm
        \epsffile{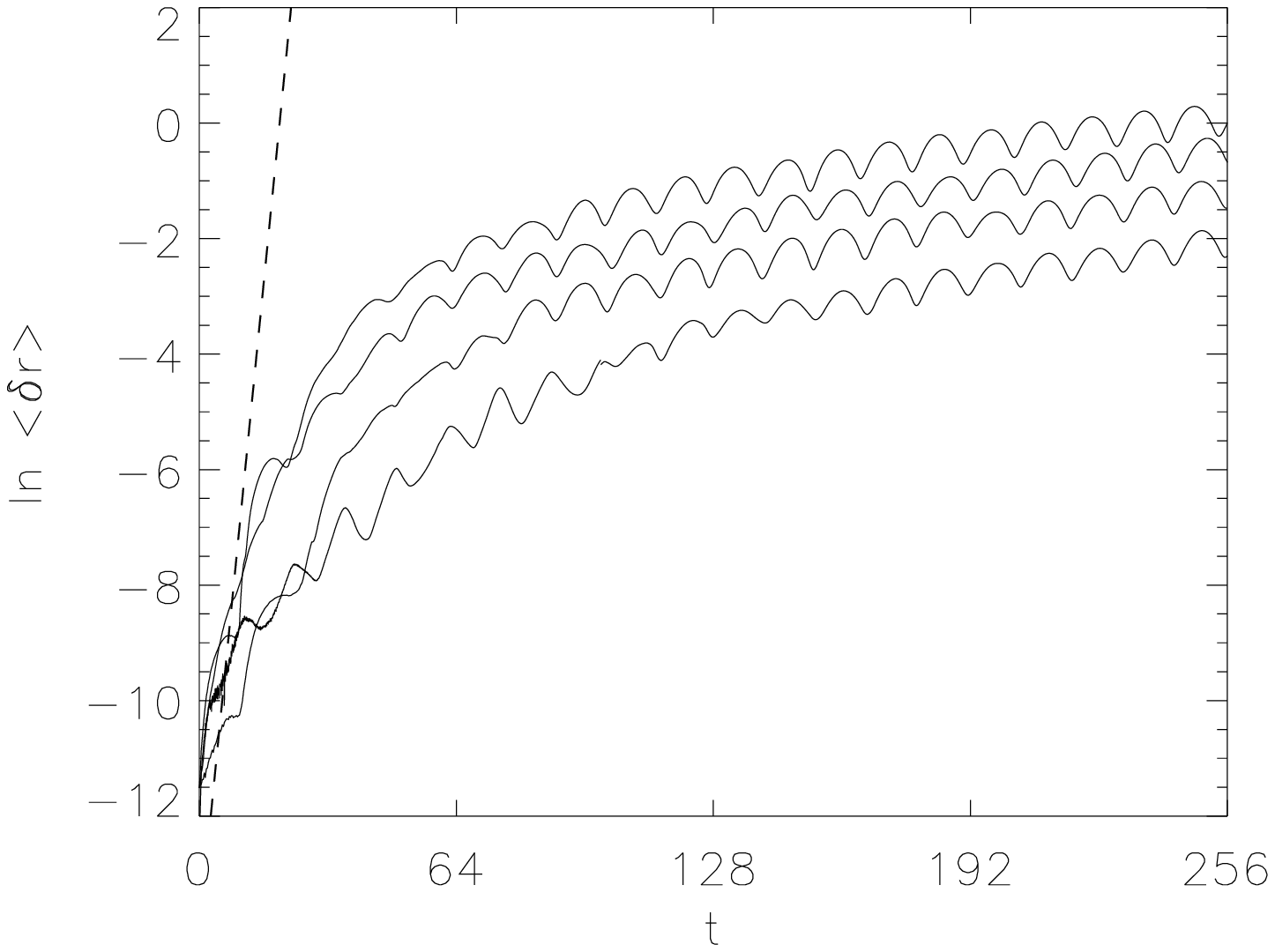}
           }
        \begin{minipage}{18cm}
        \end{minipage}
        \vskip -0.2in\hskip -0.1in
\caption{
The same as the preceding Figure, now computed for orbits in frozen-$N$
realisations of the integrable Plummer density distribution.\label{fig2}}
\vspace{0.1in}
\end{figure}

\pagestyle{empty}
\begin{figure}[t]
\centering
\centerline{
        \epsfxsize=18cm
        \epsffile{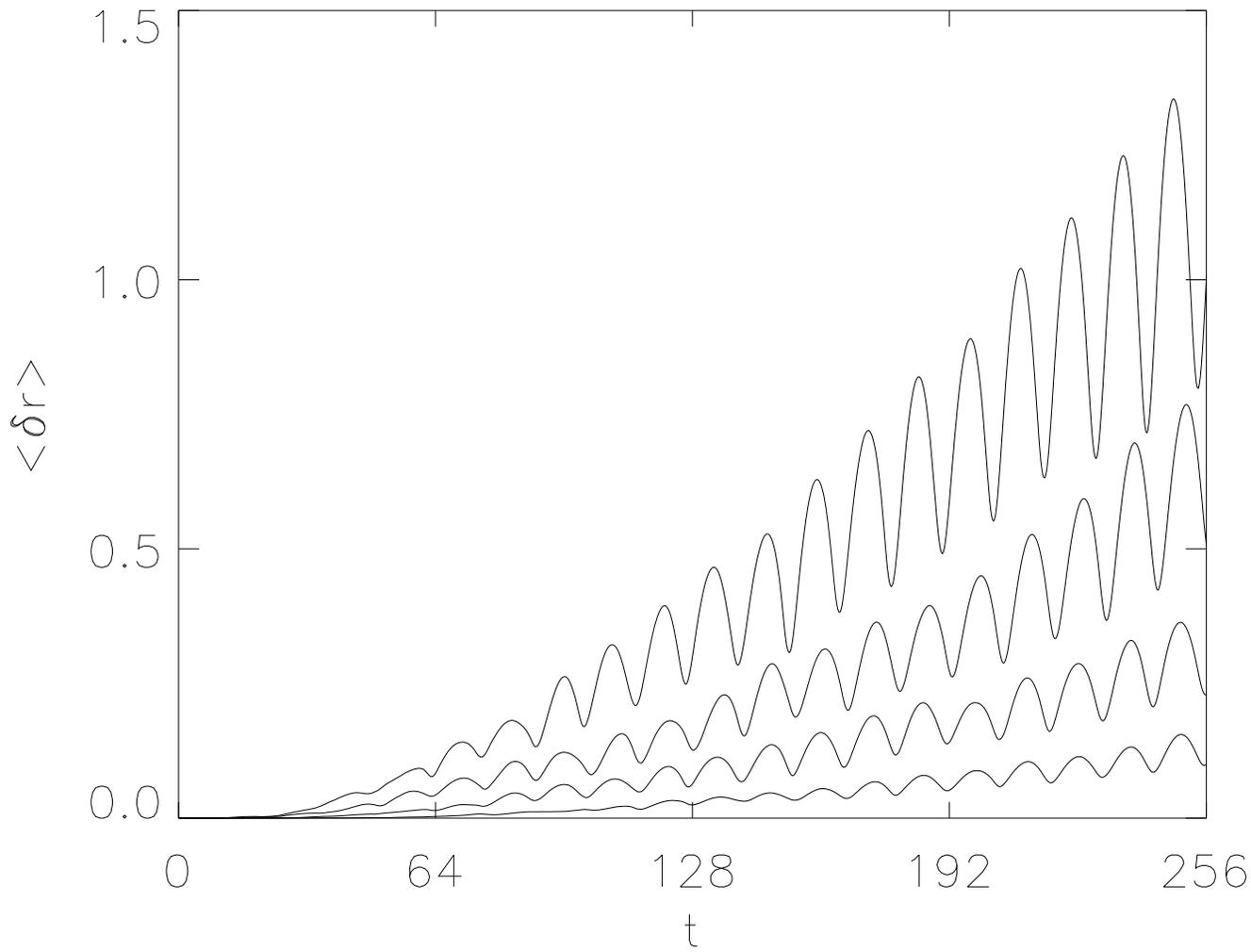}
           }
        \begin{minipage}{18cm}
        \end{minipage}
        \vskip -0.2in\hskip -0.1in
\caption{
The same data as Figure 2, now plotted on a linear scale.\label{fig3}}
\end{figure}

\pagestyle{empty}
\begin{figure}[t]
\centering
\centerline{
        \epsfxsize=18cm
        \epsffile{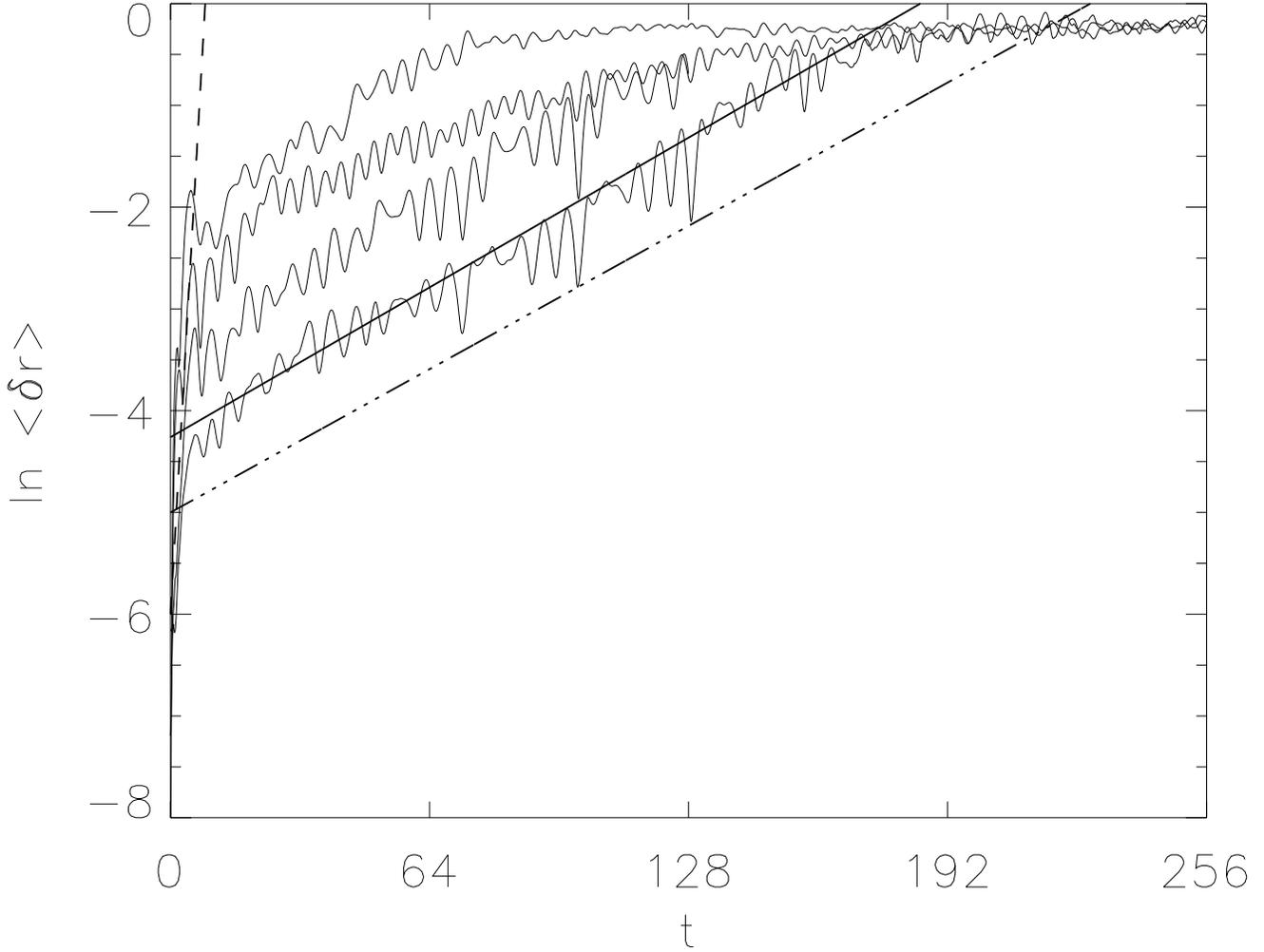}
           }
        \begin{minipage}{18cm}
        \end{minipage}
        \vskip -0.2in\hskip -0.1in
\caption{
The mean spatial separation between orbits with identical initial conditions 
evolved in two different frozen-$N$ backgrounds, each sampling the chaotic
ellipsoid plus black hole potential. As in the preceding Figures, each curve
averages over $100$ pairs of orbits. The four curves correspond (from top to 
bottom) to $N=10^{4.5}$, $10^{5}$, $10^{5.5}$, and $10^{6}$. The solid line 
corresponds to a slope of $0.023$, generated as a least squares fit to the
$N=10^{6}$ data over the interval $32<t<160$. The triple dot-dashed line has 
a slope
$0.022$, equal to the mean value of the smooth potential Lyapunov exponent
${\chi}_{S}$. The dashed curve has a slope $0.75$, equal to the mean value of 
the $N$-body Lyapunov exponent ${\chi}_{N}$.\label{fig4}}
\vspace{0.1in}
\end{figure}

\pagestyle{empty}
\begin{figure}[t]
\centering
\centerline{
        \epsfxsize=18cm
        \epsffile{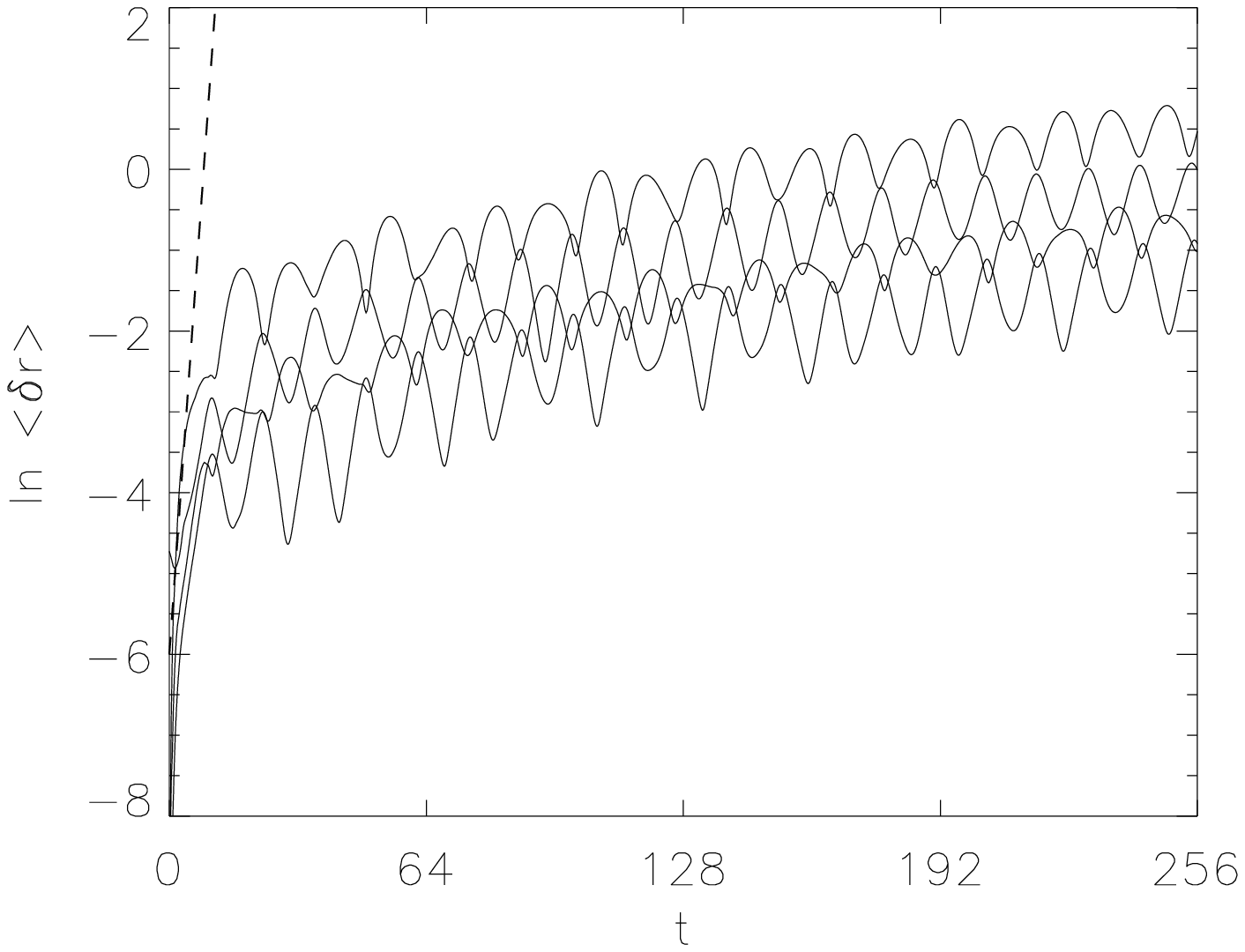}
           }
        \begin{minipage}{18cm}
        \end{minipage}
        \vskip -0.2in\hskip -0.1in
\caption{
The same as the preceding Figure, now computed for orbits in frozen-$N$
realisations of the integrable Plummer density distribution.\label{fig5}}
\vspace{0.1in}
\end{figure}

\pagestyle{empty}
\begin{figure}[t]
\centering
\centerline{
        \epsfxsize=18cm
        \epsffile{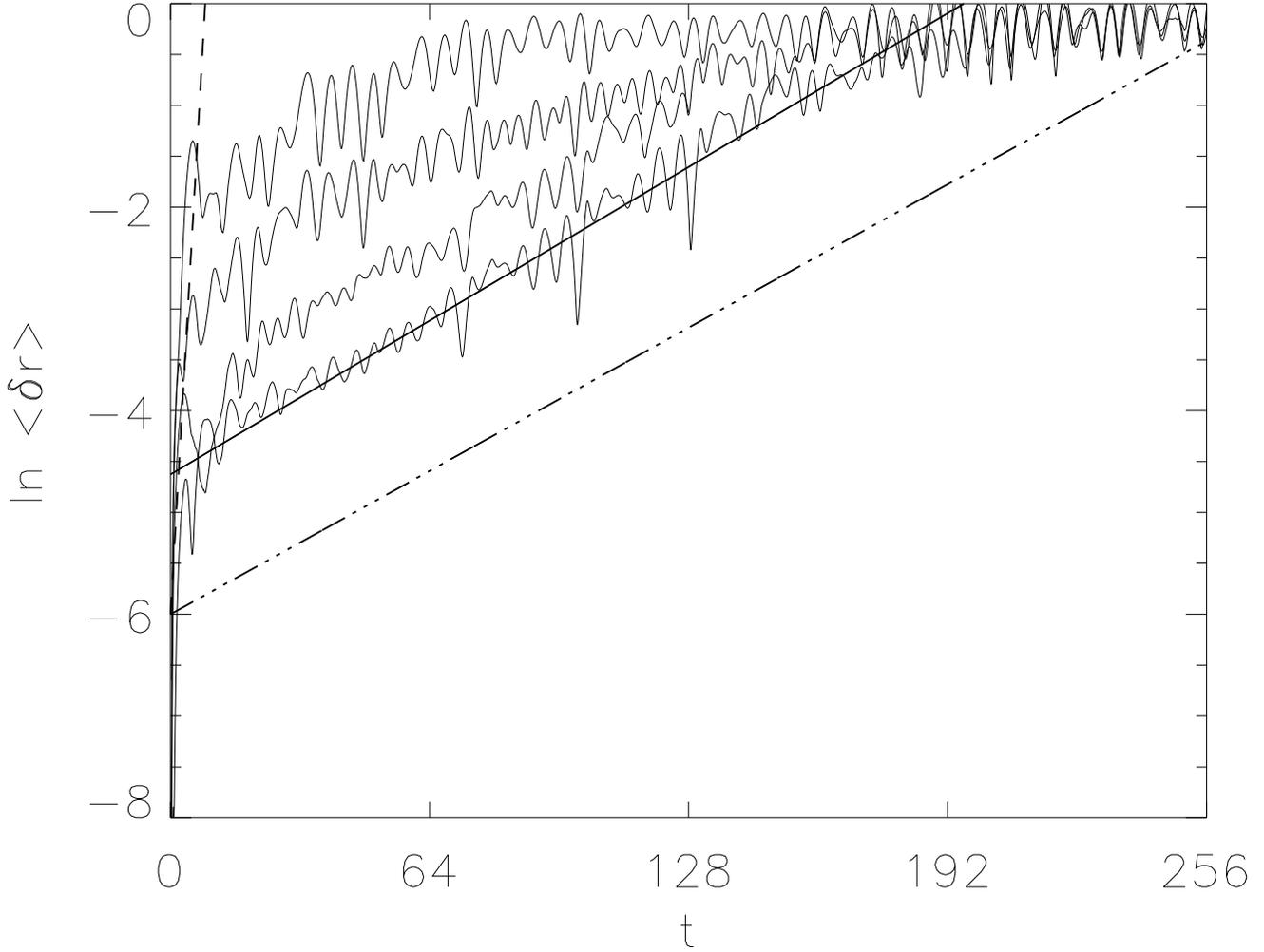}
           }
        \begin{minipage}{18cm}
        \end{minipage}
        \vskip -0.2in\hskip -0.1in
\caption{
The mean spatial separation between orbits with identical initial conditions
evolved in the smooth ellipsoid plus black hole potential and a frozen-$N$ 
realisation thereof. As in the preceding Figures, each curve averages over
100 initial conditions. The four curves again correspond (from top to bottom) 
to $N=10^{4.5}$, $10^{5}$, $10^{5.5}$, and $10^{6}$. The solid line 
corresponds to a slope of $0.023$, generated as a least squares fit to the
$N=10^{6}$ data over the interval $32<t<160$. The triple dot-dashed line 
has a slope $0.022$, equal to the mean value of the smooth potential Lyapunov 
exponent ${\chi}_{S}$. The dashed curve has a slope $0.75$, equal to the mean 
value of the $N$-body Lyapunov exponent ${\chi}_{N}$.\label{fig6}}
\vspace{0.1in}
\end{figure}

\pagestyle{empty}
\begin{figure}[t]
\centering
\centerline{
        \epsfxsize=18cm
        \epsffile{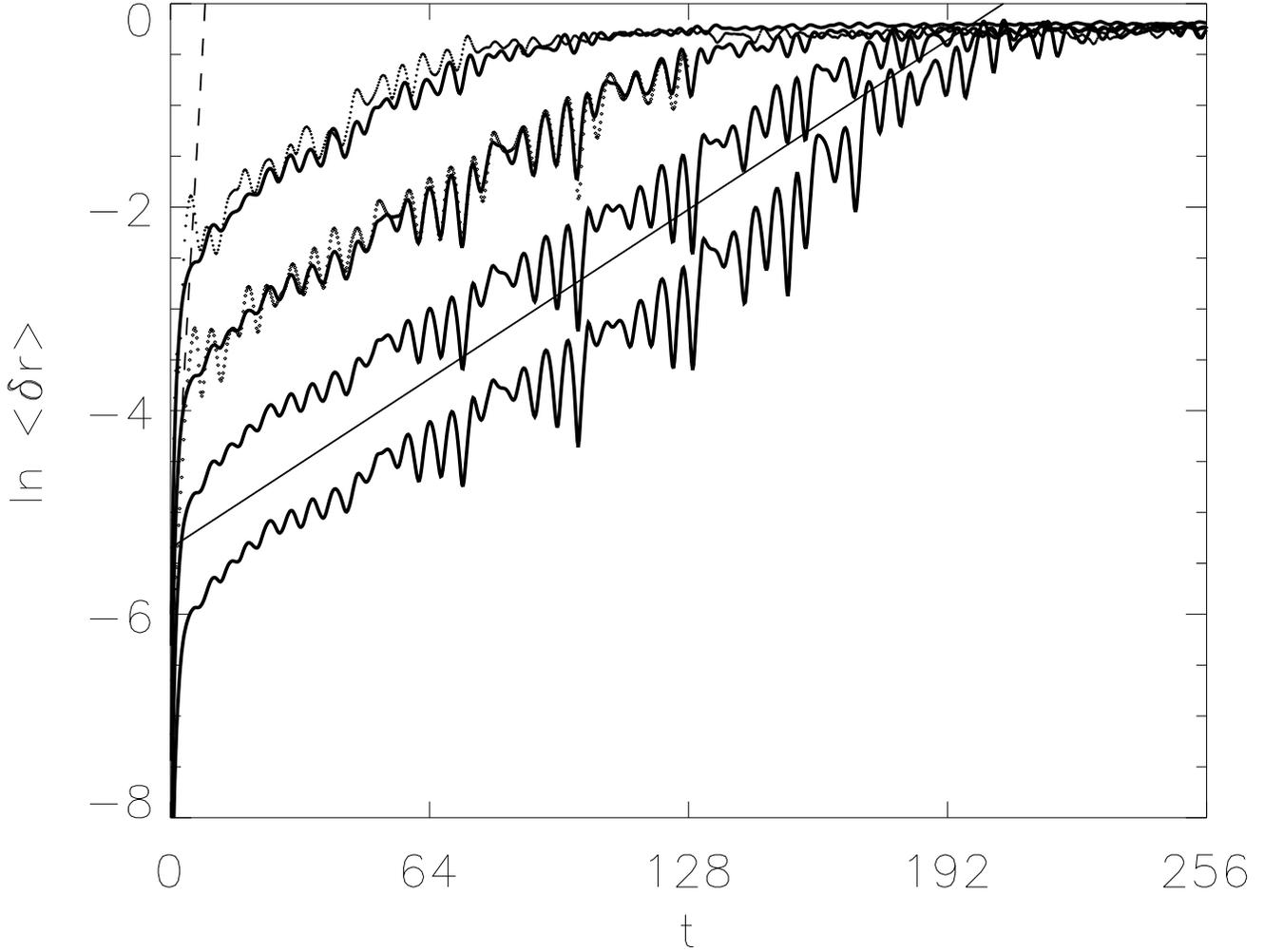}
           }
        \begin{minipage}{18cm}
        \end{minipage}
        \vskip -0.2in\hskip -0.1in
\caption{
The mean spatial separation between orbits with identical initial conditions
evolved in the smooth potential in the absence of perturbations and in the
presence of energy-conserving white noise. The different
curves correspond (from top to bottom) $\log_{10}\eta=-4,-5,-6$ and $-7$.
The solid line has a slope
$0.022$, equal to the mean value of the smooth potential Lyapunov exponent
${\chi}_{S}$. The dashed curve has a slope $0.75$, equal to the mean value of 
the $N$-body Lyapunov exponent ${\chi}_{N}$.
The dotted lines reproduce the curves in Figure 4 for $N=10^{4.5}$ and 
$10^{5.5}$, corroborating the claim that discreteness effects for 
$N=10^{p+1/2}$ are well-mimicked by ${\eta}=10^{-p}$.
\label{fig7}}
\vspace{0.1in}
\end{figure}

\end{document}